# SOME PRACTICAL ISSUES OF THE TRACKING PROCESS IN GNSS RECEIVERS

*Shafran S.V., Kudryavtsev I.A., Kumarin A.A.*
*Samara University, Samara, Russian Federation*
*E - mail: mailbox-kddk@mail.ru*

The accuracy and noise immunity of GNSS receivers are largely determined by the performance of their tracking modules. To optimize performance, designers should take into account several key issues: the choice of integration intervals, the time delay (spacing) between the early and late correlator channels, and the parameters of tracking loop filters. This article presents the results of an analysis of these important practical aspects. The obtained results display some ways of improving the parameters of transient processes, which occur in tracking modules. The study was conducted on an experimental navigation receiver, processing GPS L1 signals. Experimental data were obtained using the simulation software package, developed by the authors, Xilinx environment Vivado, as well as the processing of recordings of real navigation signals obtained in full-scale tests.

## Introduction

Tracking units for navigation receivers have been described in many papers, such as [1–3]; however some practical aspects have not been covered in sufficient detail. For example, receivers operating with weak signals utilize an extended integration interval; in conditions with significant reflected signals (multipath), reduced delays between the late and early (E-L) channels are applied [4]. Some authors [5, 6] investigated the problem of tracking accuracy and provided recommendations for selecting the tracking loop parameters. Papers [7, 8] consider practical aspects of filter implementation. The present paper discusses some specific issues arising in practical implementation.

One of the common solutions is using the popular chip RF front-end MAX2769 [9], configured to output in-phase/quadrature (I/Q) signal components at a sampling rate of 16.368 MHz. MAX2769 provides automatic gain control (AGC) and performs filtering and downconversion to a selected intermediate frequency. The navigation signal acquisition process is implemented in software using an algorithm providing initial estimation of the Doppler frequency shift with an error of ±100 Hz. Ranging code delay was estimated with an accuracy of about 10–15 reference clock cycles.

Tracking modules, monitoring navigation signals, are usually implemented in accordance to the classical architecture shown in Figure 1 [4].

Some advantages in reducing errors in the code channel can be achieved by using a modified shift register with adjustable delay between early and late channels. The receiver implementation includes several FPGA-based tracking modules performing continuous signal monitoring under the control of a processor core (CPU) responsible for module configuration and position computation. The modules generate interrupts at the end of every integration interval, and the CPU receives the data and performs necessary calculations, using floating-point arithmetic. The required configuration parameters (code and carrier NCO increments) are loaded back into the modules.

The simulation program developed in this work generates a simplified version of the GPS navigation signal with adjustable parameters: carrier phase shift and Doppler frequency, as well as code delay. The simulator can insert phase transitions corresponding to navigation data bits and add white Gaussian noise (obtained using pseudo-random generator) to achieve a target carrier-to-noise density ratio ($C/N_0$). The program was used to investigate the behavior of tracking modules under given input signal characteristics. The main goal of the simulation was to test various tracking loop configurations. The filters used in these loops are described in detail in [2, 3]. As a rule, developers are interested in reliable signal acquisition and accurate tracking. Accuracy of the code and carrier tracking can be estimated by the evaluation of phase/frequency deviations of the local NCOs from the parameters of received signal.

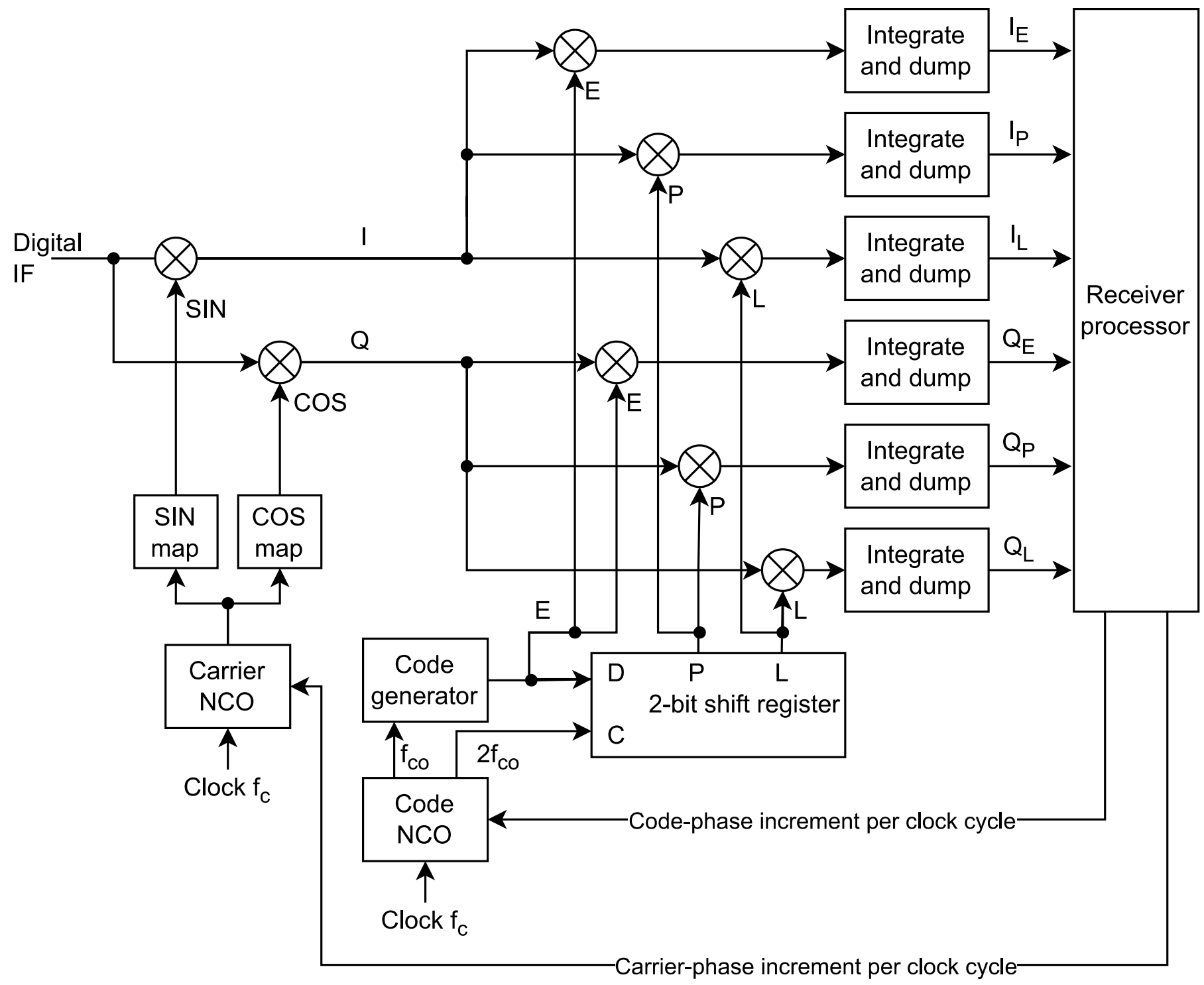

Figure 1. Architecture of the correlation channel in a common navigation receiver [4]

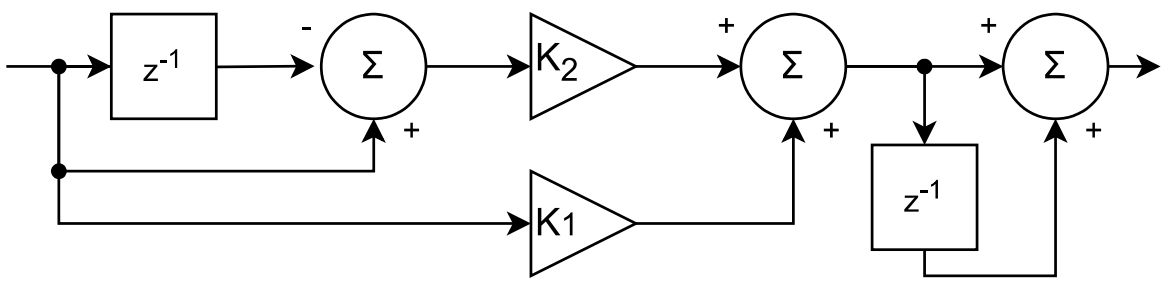

Figure 2. Block diagram of a Type I filter

This work compares second-order filters [2,3] with the third-order filter from [2]. A special place is occupied by a filter based on the Matlab implementation of the navigation receiver provided in the electronic appendix of [3]. The block diagram of this filter (hereinafter, the Type I filter) is shown in Figure 2. The circuit diagrams of the filters from [2] (Type II - second-order filter, Type III - third-order filter) are shown in Figures 3a and 3b.

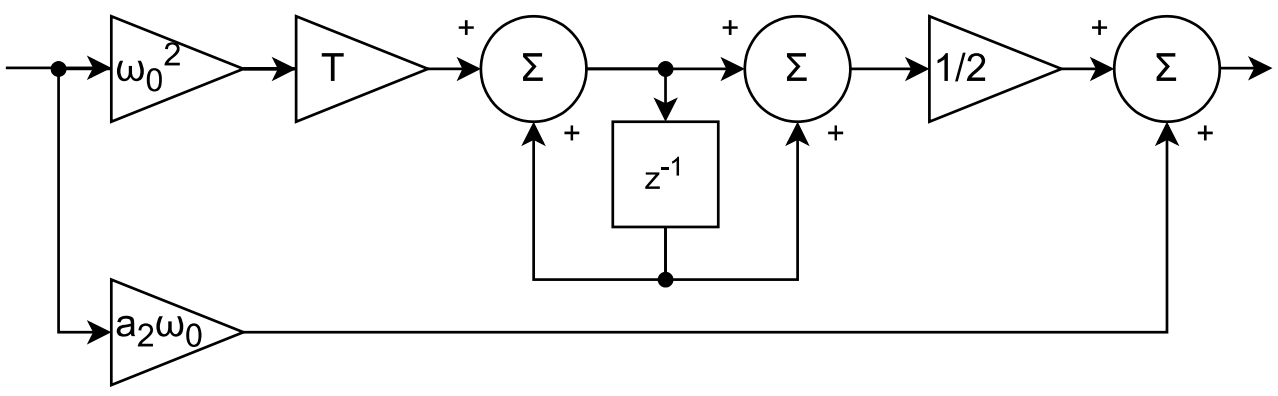

(a) second order filter

The filter parameters are set according to Table 1, where $\omega_0$ is the frequency (Hz), $B_n$ is the noise bandwidth (Hz), $\zeta$ is the damping coefficient, and T is the integration interval (s). The description of the parameters is provided in exact accordance with the sources [2,3].

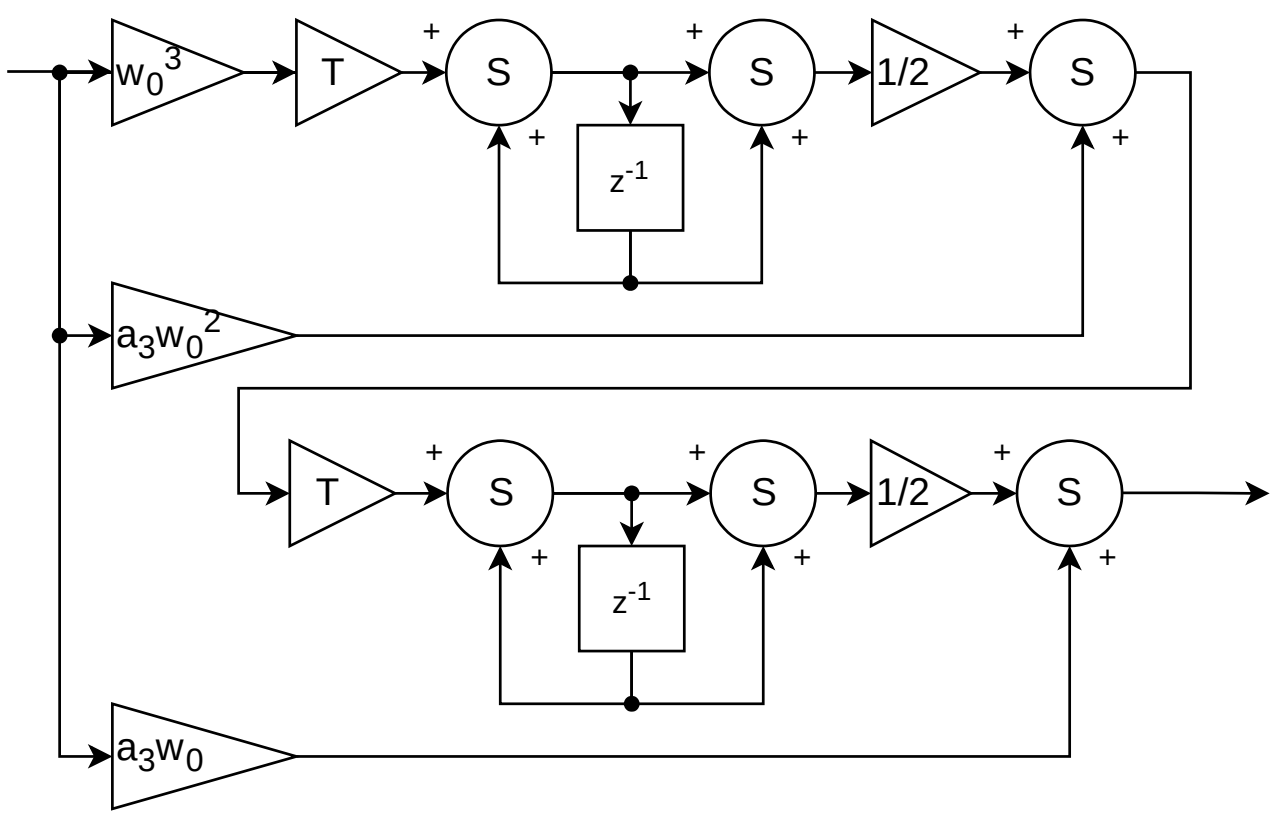

(b) third order filter

Figure 3. Block diagrams of the filters proposed in [4]

Table 1. Filter parameters

| Filter | Parameters |
|---|---|
| Type I | $B_n = \frac{\omega_0(4\zeta^2+1)}{8\zeta}$, $K_1 = 2\zeta\omega_0$, $K_2 = \omega_0^2 T$ |
| Type II | $B_n = \frac{\omega_0(1+a_2^2)}{4a_2}$, $a_2 = 1{,}1414$, $B_n = 0{,}53\omega_0$ |
| Type III | $B_n = \frac{\omega_0(a_3 b_3^2 + a_3^2 - b_3)}{4(a_3 b_3 - 1)}$, $a_3 = 1{,}1$, $b_3 = 2{,}4$, $B_n = 0{,}7845\omega_0$ |

### Using extended integration intervals

Employing increased integration intervals is beneficial when working with weak signals. It also reduces the tracking error and the number of processed interrupts, used in software defined receivers. At the same time, in receivers processing signals containing navigation data, the integration interval cannot exceed the data bit length. For the classic GPS L1 signal, this limit is 20 ms, whereas for GLONASS L1, it is 10 ms, due to the specific structure of the navigation message.

To align the integration intervals with the data bit boundaries, the tracking module in our experiment starts with a 1-ms interval and then switches to an extended period after detecting the phase transition caused by the transmission of a navigation bit. With such an increase in the integration interval, desynchronization can occur due to an erroneous phase transition detection. Lock detectors, similar to those described in [3], can fail to detect such an error.

Figure 4 shows the output signal of an extended interval integrator (in arbitrary units), reset to zero at the beginning of each integration interval. Under normal conditions, the signal within the integration interval is more or less monotonic. However, if misaligned with the navigation data, a phase transition occurs leading to a sign change of the signal phase and consequently, to non-monotonic behavior within the integration interval.

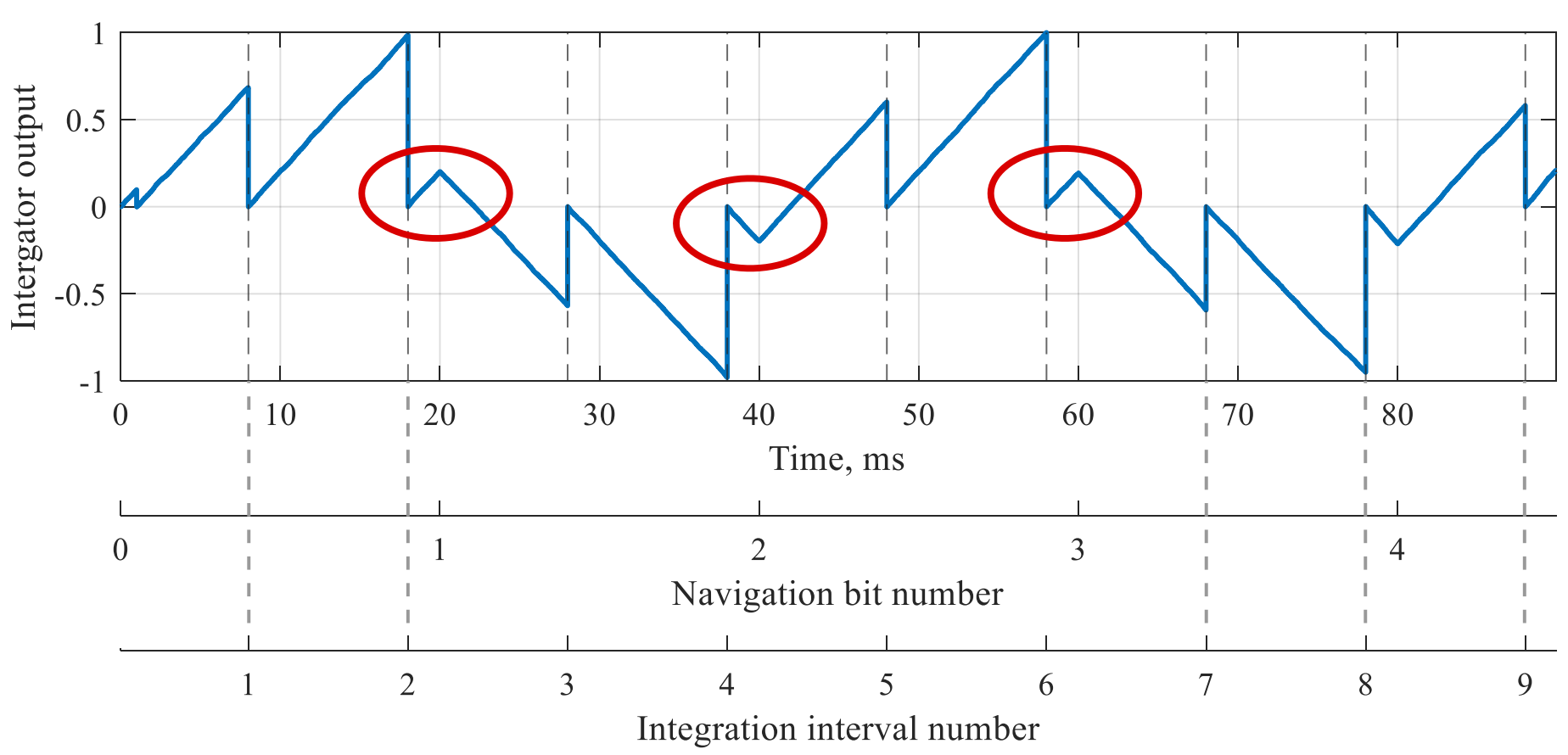

Figure 4. Tracking error for a 10-ms integration interval

To prevent this misalignment, the tracking module must determine the data bit boundary in advance. Since the phase transitions caused by the navigation bits coincide with the epoch boundaries, a sign change in the integrator's output indicates a data bit boundary. After detection of a bit boundary, the module must integrate for (N-1) ms (switching to N ms integration from 1-ms mode); subsequently the interval boundaries will be aligned with the bit boundaries.

The criteria for selecting tracking loop parameters are described in detail in [5]. The main idea is to choose a filter with the narrowest possible passband, taking into account the carrier-to noise density ratio ($C/N_0$) and the dynamics of the platform on which the receiver is installed. Generally, signals with $C/N_0$ below 30 dB-Hz do not provide stable operation of tracking loops [10].

Simulation results of the tracking process for three types of filters for a stationary receiver are shown in Figure 5. The experiment was conducted under ideal conditions: the initial frequencies of the carrier and code NCOs of the tracking module match the simulated signal. The first 2 seconds used 1 ms integration, followed by a switch to 10-ms mode. A PLL bandwidth of 5 Hz was used in both modes, and the DLL loop bandwidth was 1 Hz. For the filter in [1], the damping factor, ζ, was set to 0.7. In all cases, a Costas quadrature loop with an arctan(I/Q) discriminator was used. It can be seen, that all three filters demonstrate similar behaviour.

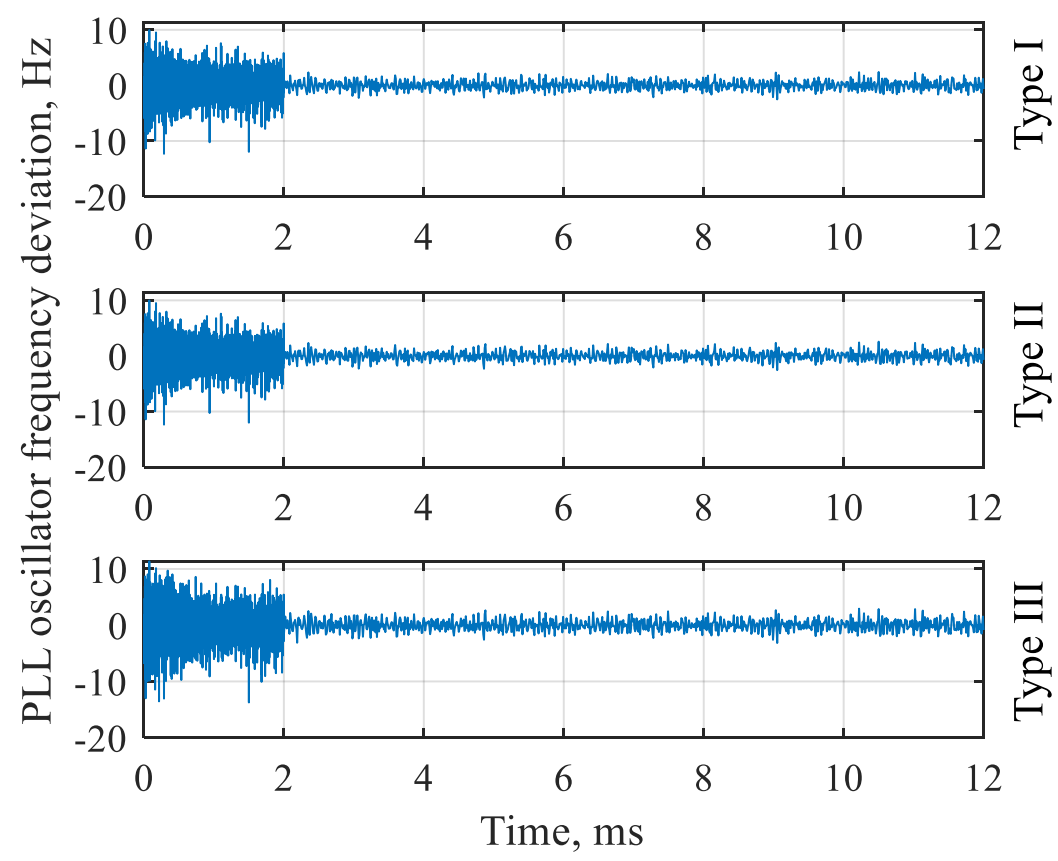

Figure 5. PLL discriminator output in $C/N_0$=39 db-Hz conditions

In the extended integration interval mode, the standard deviation (*RMS*) of the frequency does not exceed 1 Hz, whereas initially it is about 2.6 Hz.

This experiment demonstrates the characteristics of the tracking modules in a steady-state mode without transient processes.

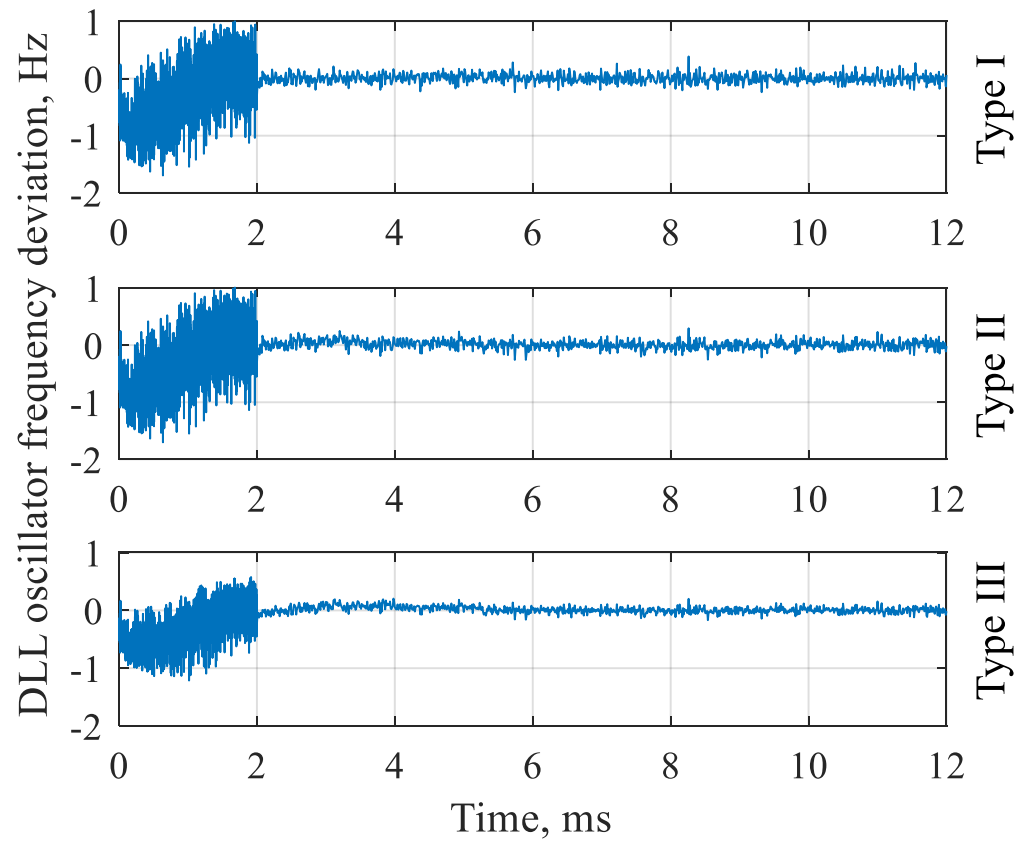

Figure 6. DLL local oscillator frequency deviation

Filters in the DLL loop with a 1 Hz bandwidth show similar results. However, the third-order filter (Type III) exhibits a lower level of fluctuations (RMS = 0.07 Hz) in the 10-ms mode, while for Types I and II filters RMS values are 0.11–0.12 Hz. A coherent dot product discriminator [2] was used in this experiment. The results are presented in Figure 6.

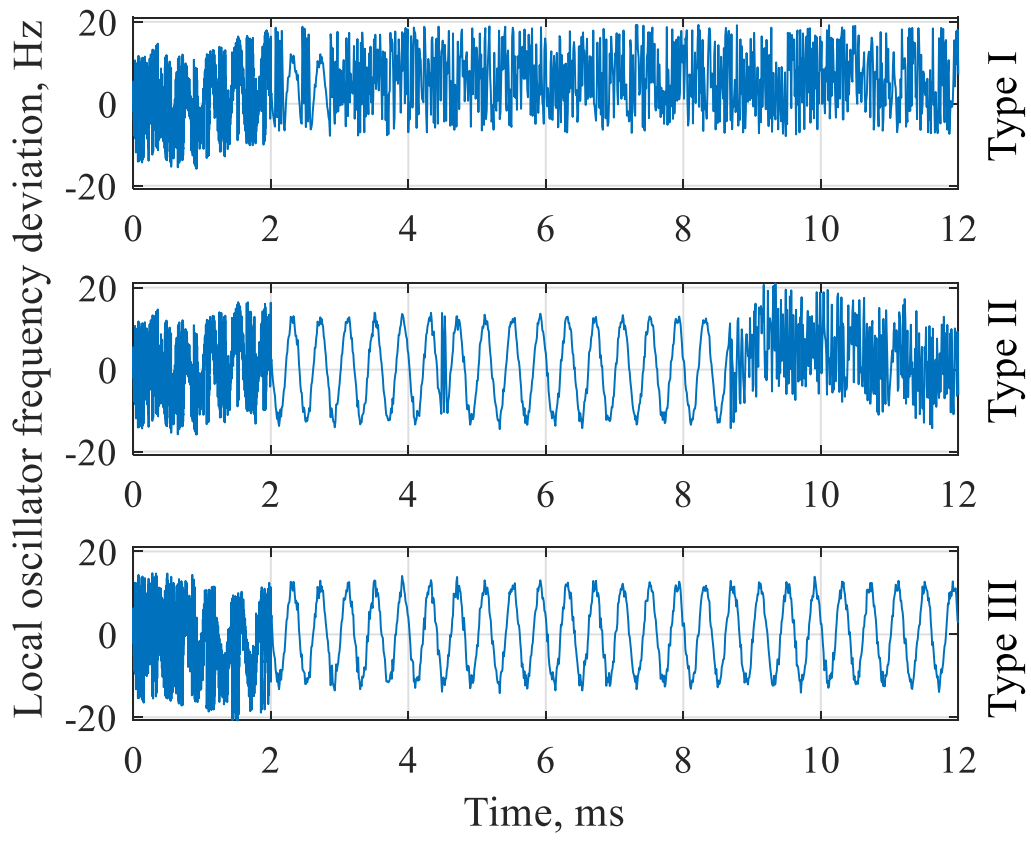

Figure 7. PLL local oscillator frequency deviation at $C/N_0$ = 35 dB-Hz

Even for a stationary receiver, the Doppler frequency and its derivative affect the tracking process. When designing a receiver for dynamic conditions, stability limitations must be taken into account. According to [2], a second-order filter has a stability margin of 30° with a bandwidth $B_n$ of about 10–20 Hz, depending on the computational delay. Although the boundary is not rigid, it should be noted that rapid changes of the Doppler frequency can cause instability. Figure 7 shows the simulation results, where the signal frequency varies harmonically with an amplitude of 11 Hz and the frequency of 2.5 Hz (PLL bandwidth was set to 5 Hz). In this experiment, the Type III filter shows significant advantages due to its better stability in dynamic conditions. Types I and II filters are less stable and lose lock under such conditions. The instability problem can be solved by increasing $B_n$. It is also evident that a 10 ms integration provides a smaller deviation of the local NCO frequency compared to the classic 1-ms option.

After signal acquisition is completed, only rough estimates of the Doppler frequency and code delay are available, and at the initial moment of tracking, the signal has an arbitrary phase. In our experiment, the Doppler frequency was estimated with a resolution of 100 Hz; therefore, we assume the error can reach 50–60 Hz. For reliable pull-in, a bandwidth $B_n$ of at least 25 Hz is required, as recommended in [3]. However, this value may exceed the stability limit in case of extended integration. The solution is to start tracking with a short interval and gradually increase its duration to the desired value as the carrietypeNCO frequency stabilizes.

In the next experiment, an initial carrier frequency offset of 40 Hz without code or phase errors was used. At a high $C/N_0$, all three filters provide a relatively fast (from the perspective of further navigation data processing) transient response, accompanied by oscillations. As can be seen in Figure 8a, the curves for filters I and II almost coincide. Evidently, the transient response duration increases with the noise level.

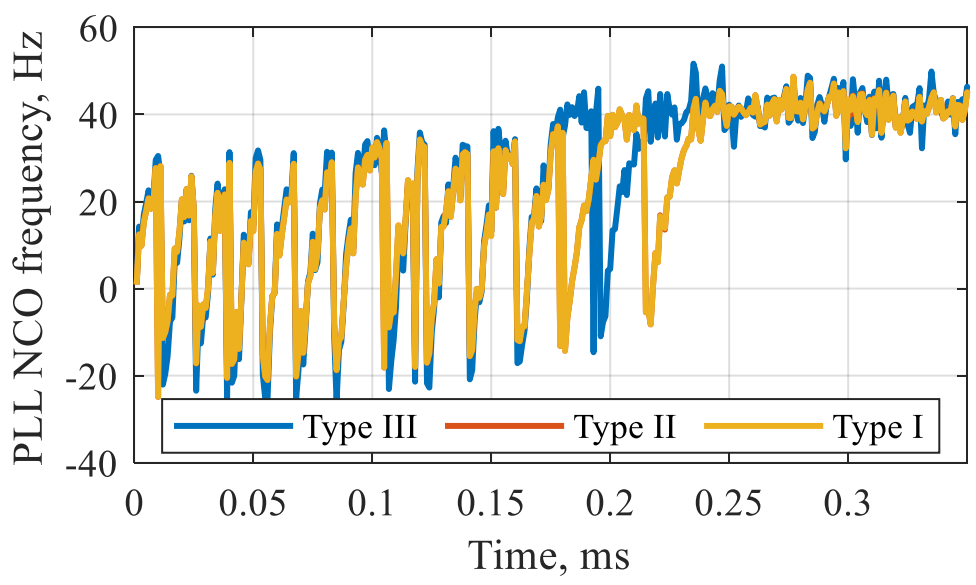

(a) $C/N_0$ = 45 dB-Hz, $B_n$ = 10 Hz

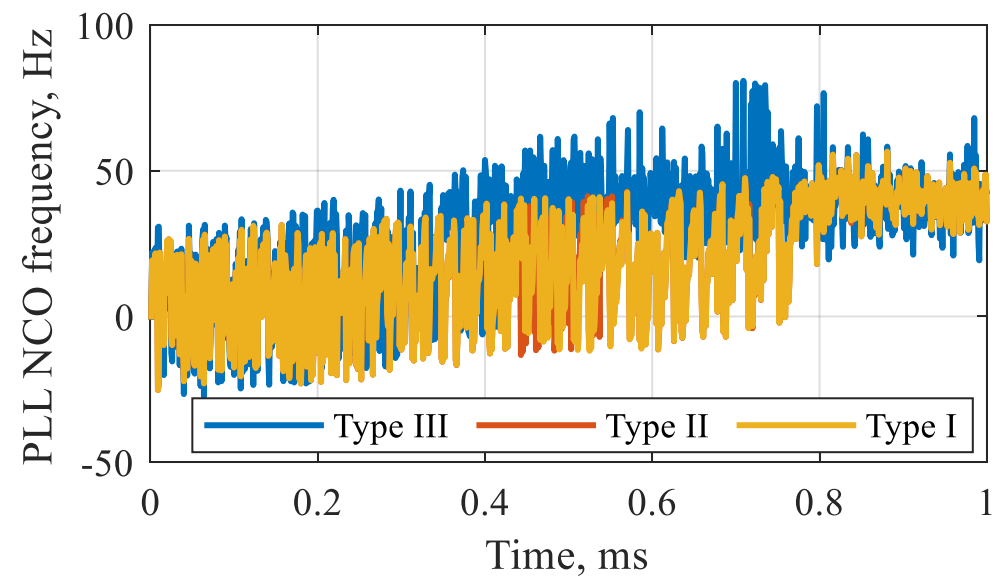

(b) $C/N_0$ = 35 dB-Hz, $B_n$ =25 Hz

Figure 8. PLL NCO frequency

Switching to an extended integration interval can lead to loss of lock. Therefore, it is advisable to use a smooth transition algorithm. A similar situation is shown in Figure 9, where various initial carrier phase shifts were simulated. In general, the sys-tem is phase-insensitive, exception for the transition from 1 ms to 10 ms for the Type I filter, which is more robust to frequency deviations.

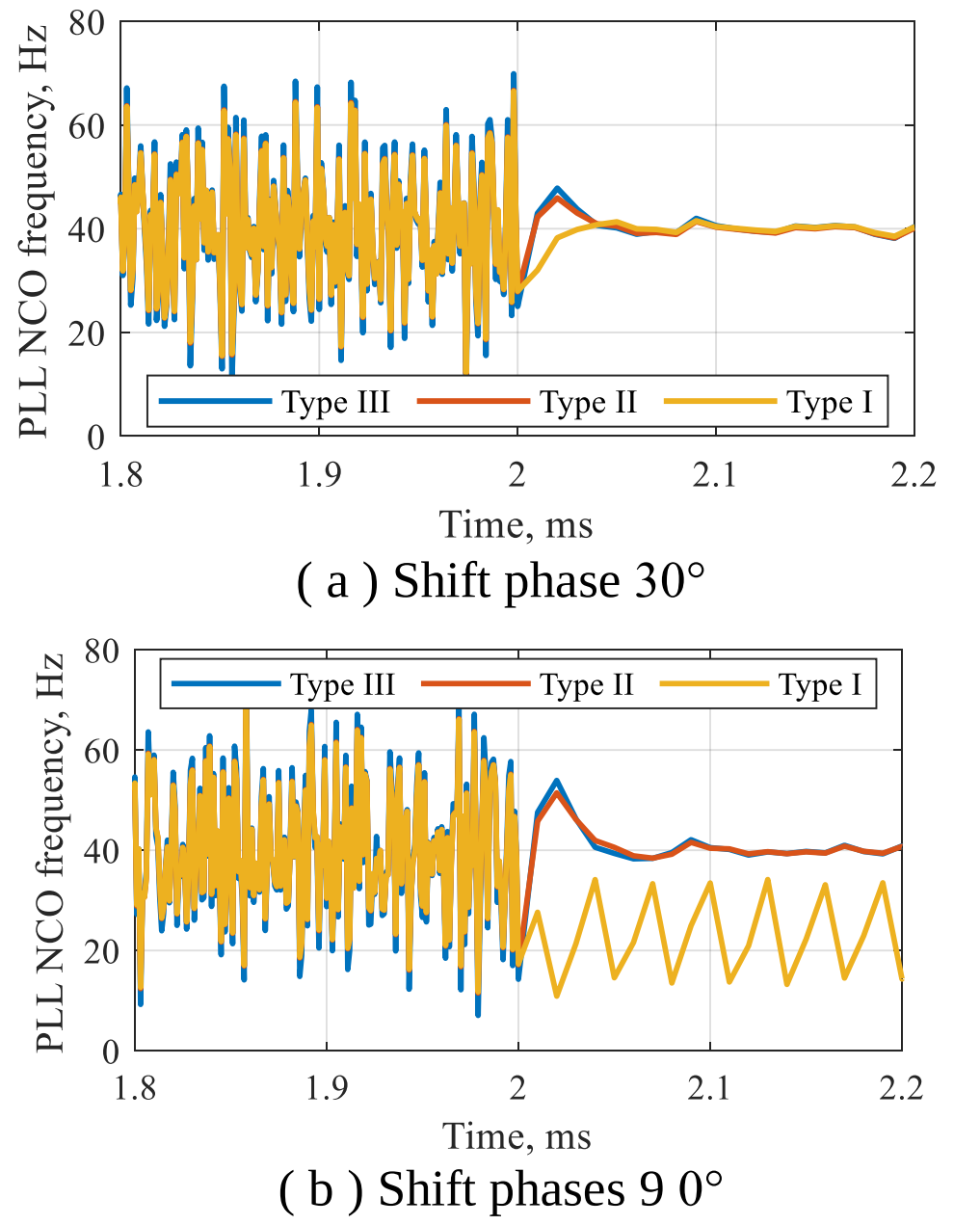

( a ) Shift phase 30°

( b ) Shift phases 9 0°

Figure 9. Effect of initial phase shift on capture quality

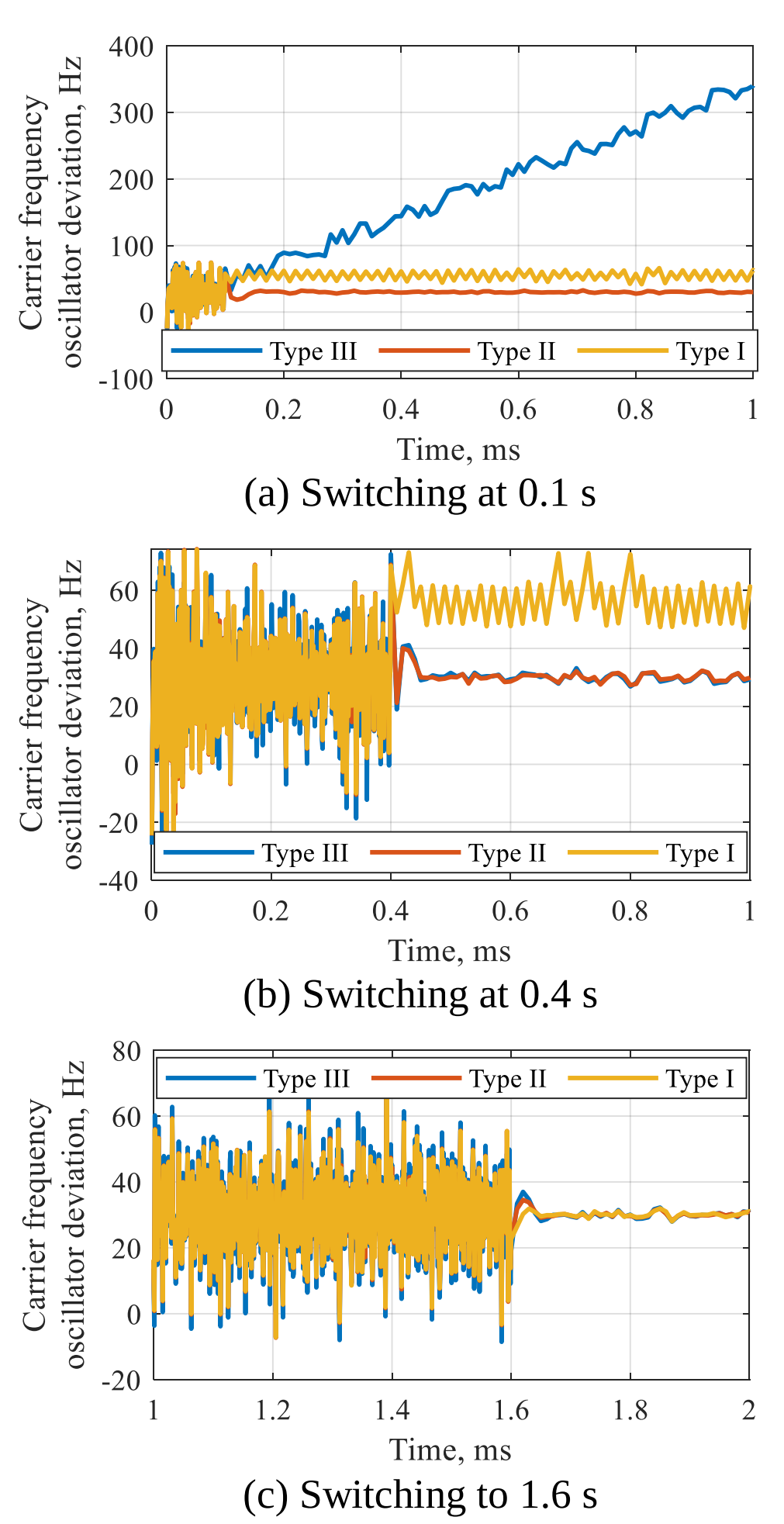

(a) Switching at 0.1 s

(b) Switching at 0.4 s

(c) Switching to 1.6 s

Figure 10. The influence of switching time on capture quality

A problem can also occur if the transition to extended integration happens too early, before the NCO frequency reaches the required value. Usually, loss of lock occurs when the oscillator frequency error exceeds the bandwidth of the filter, as shown in Figure 10.

This transition also affects the lock detector metric. The Type I filter is particularly sensitive to this stress. Using this metric without additional measures can result in a false loss of tracking. The simulation results are shown in Figure 11. The long transient response is partly explained by signal averaging.

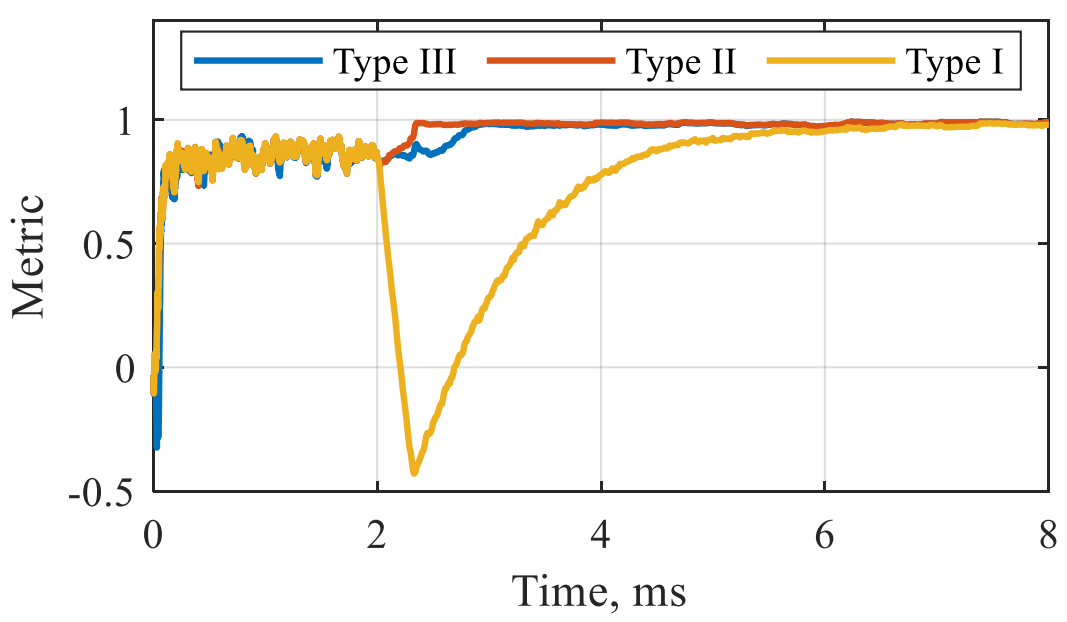

Figure 11. Lock detector metric

When choosing a discriminator for a DLL, it is important to consider the operating mode. At the initial stage of tracking, it is advisable to use a non-coherent discriminator, and after PLL lock - a coherent one (scalar product discriminator), which is easier to implement [2].

Figure 12 shows the simulation results of the DLL NCO frequency behaviour. In all cases, initial delay errors were not introduced; however, the modules have an inherent offset due to the specifics of the internal shift register implementation. At a $C/N_0$ exceeding 35 dB-Hz, the Type I filter loses lock when switching to extended integration (not shown), as discussed above.

Figure 13 shows the transient process in the DLL for various initial code delays (a clock cycle corresponds to the sampling interval) at $C/N_0$ = 40 dB-Hz. In cases (a) and (b), the results for Type I and III filters are almost identical. It can be noted that the frequency deviation of the DLL generator in the 1-ms mode and the duration of the transient process depend on the code delay value. With a delay of 20 samples, the transient process of the Type II filter does not complete within 2 seconds.

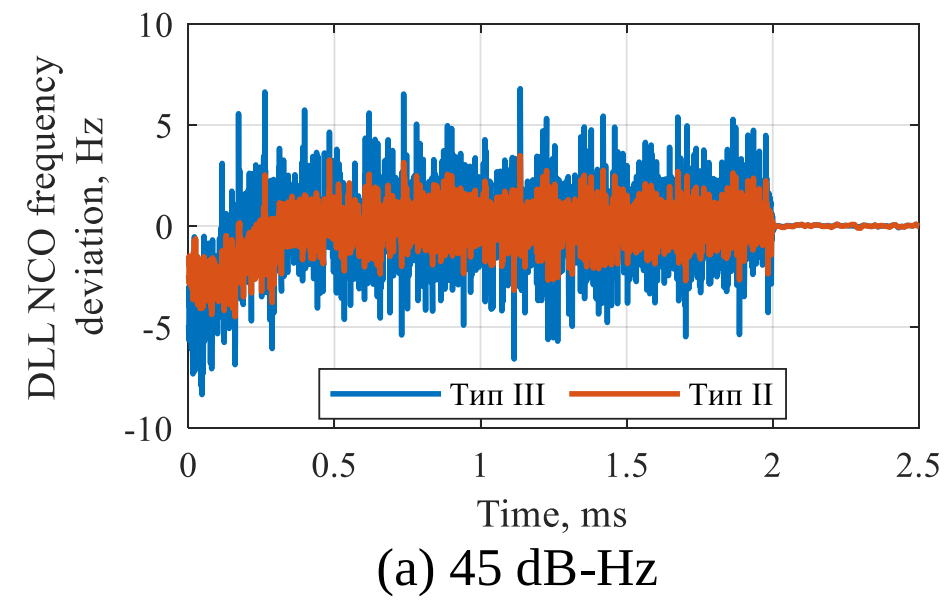

(a) 45 dB-Hz

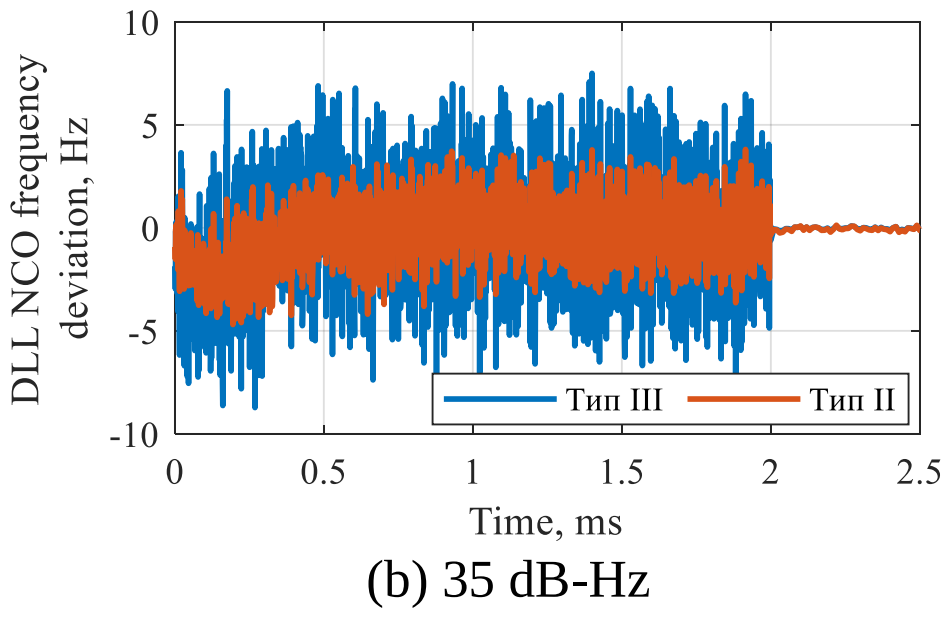

(b) 35 dB-Hz

Figure 12. NCO frequency deviation for different values of $C/N_0$ . $B_n = 25$ Hz, $B_n = 5$ Hz. The curves for filters of Type I and II are the same.

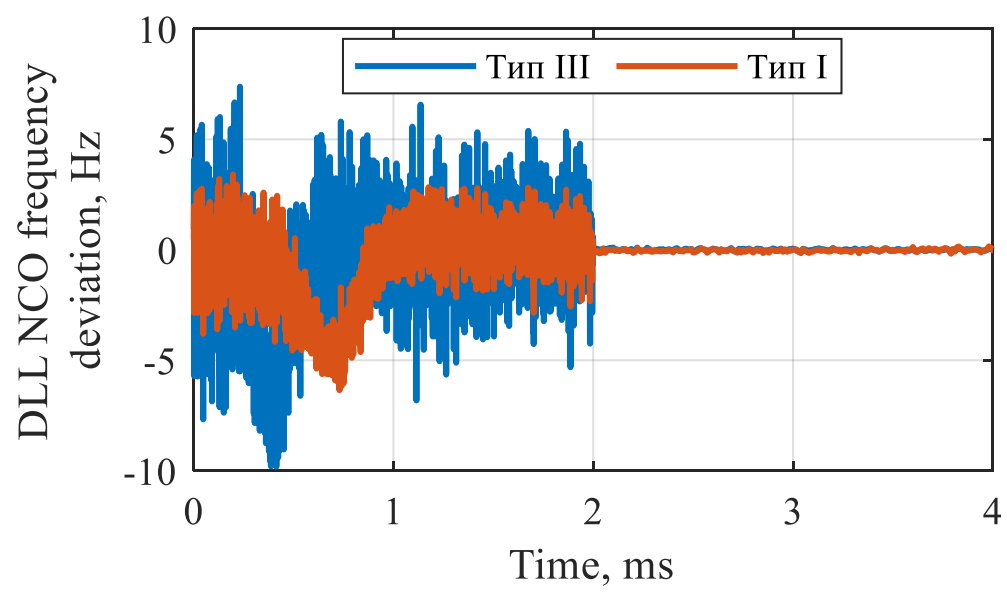

(a) Code delay is 15 cycles. The curves for filter Types I and II are the same.

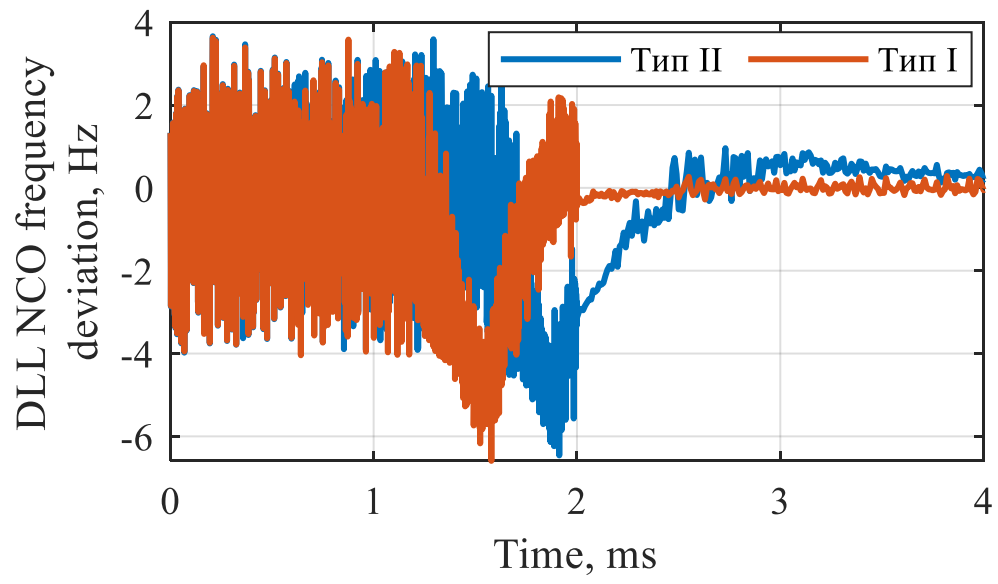

(b) Code delay is 20 cycles. The curve for the Type II filter is intentionally shifted upward by 10 Hz. The Type III filter was not operational from the very beginning of tracking.

Figure 13. Deviation of the generator frequency in the DLL circuit from the nominal value of 1.023 MHz for different initial code delays

## Using Reduced Spacing in the Code Tracking Loop

The specifics of applying short delays between the early, prompt, and late channels are described in detail in [4]. This approach is useful in multipath conditions and provides a lower tracking error. Typically, the spacing is 0.5 chips (code sequence elements); in the short-range mode, it can be reduced to 0.1 chips, and sometimes to 0.05 chips. According to [4], the RMS frequency of the local oscillator frequency is $\sigma^2 \sim d$, where d is the channel spacing. However, transitioning to a small spacing can cause a longer transient process in the tracking loop. An example of such a process is shown in Figure 14.

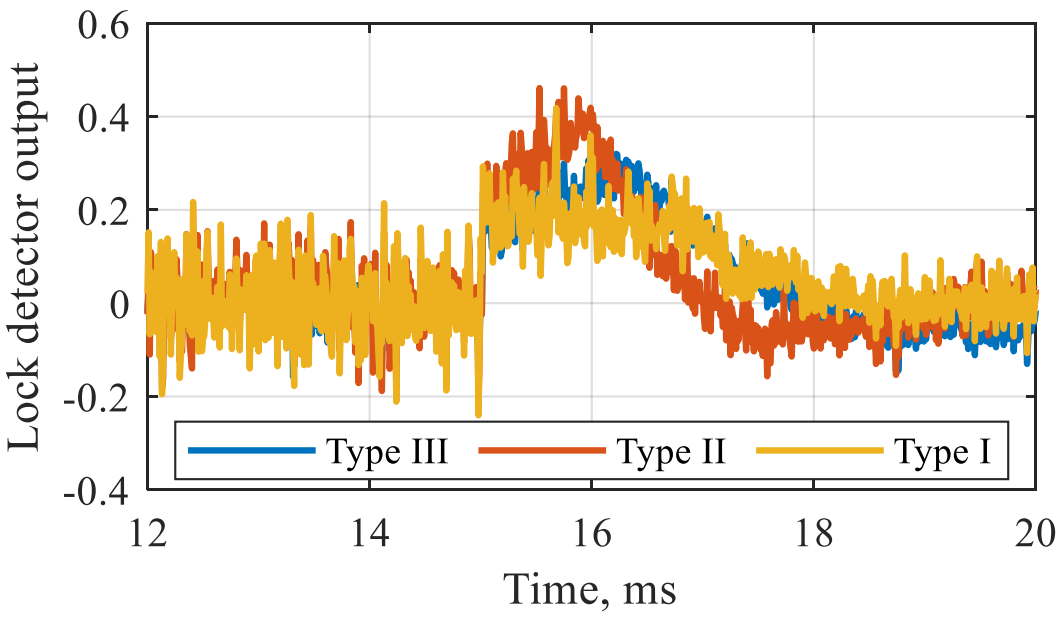

( a ) Lock detector output

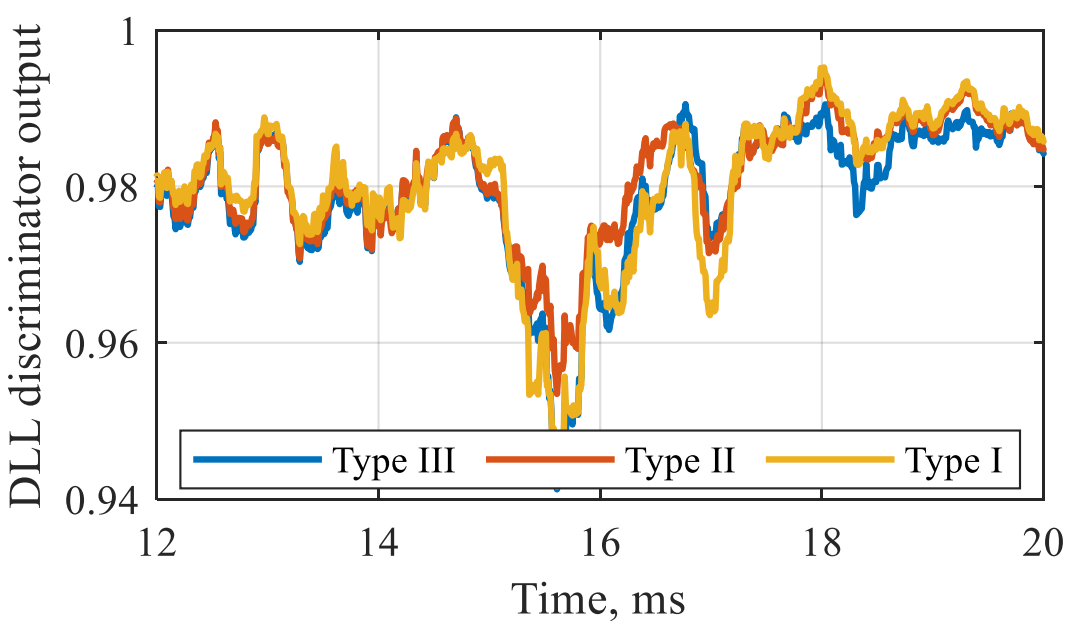

(b) Output signal of the discriminator in the DLL tracking loop

Figure 14. Transient process after a decrease in spacing

Although this transient does not create a serious problem and can be ignored, with a narrow DLL bandwidth, it can increase the total error for several seconds after switching. If necessary, a stepwise reduction in spacing can be implemented. In this case, the DLL frequency deviation decreases. RMS measurements show that by reducing spacing from 0.5 to 0.1 chip, it is possible to reduce the RMS deviation from approximately 0.07 to 0.036 for a Type II filter (at $C/N_0 = 39$ dB-Hz ).

## Tracking a Real GPS Signal

The results presented below were obtained from a GPS receiver using an RF front-end based on the MAX2769, with the software acquisition module and hardware tracking modules implemented on an FPGA. During the experiment, the $C/N_0$ level of the input signal was about 44 dB-Hz. The results of tracking one of the satellites for 10 seconds from the moment of acquisition are shown in Figures 15–17. Tracking started with 1 ms integration; then, after about 2 s, the integration interval was increased to 10 ms. The PLL bandwidth was initially 25 Hz and was reduced to 5 Hz in extended integration. The DLL bandwidths were 3 Hz and 0.8 Hz, respectively. A second-order filter (Type II) was implemented. The correlator channel spacing was initially 0.5 chips, and then was then gradually reduced to about 0.1 chips (approximately at the $4^{th}$, $6^{th}$, and subsequent seconds).

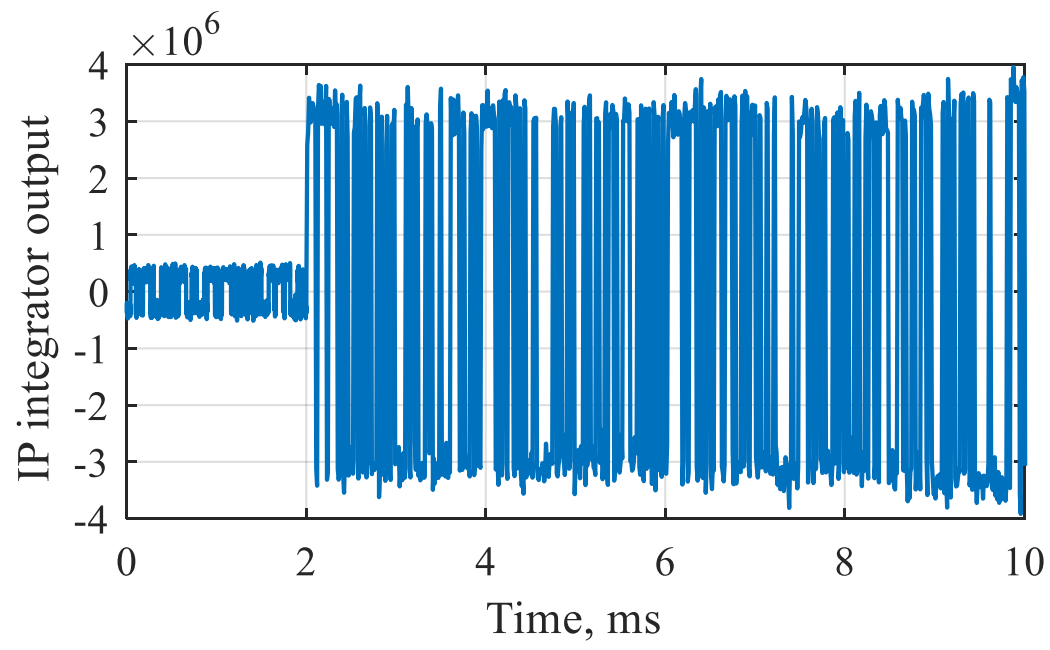

Figure 15. IP channel integrator output (prompt signal)

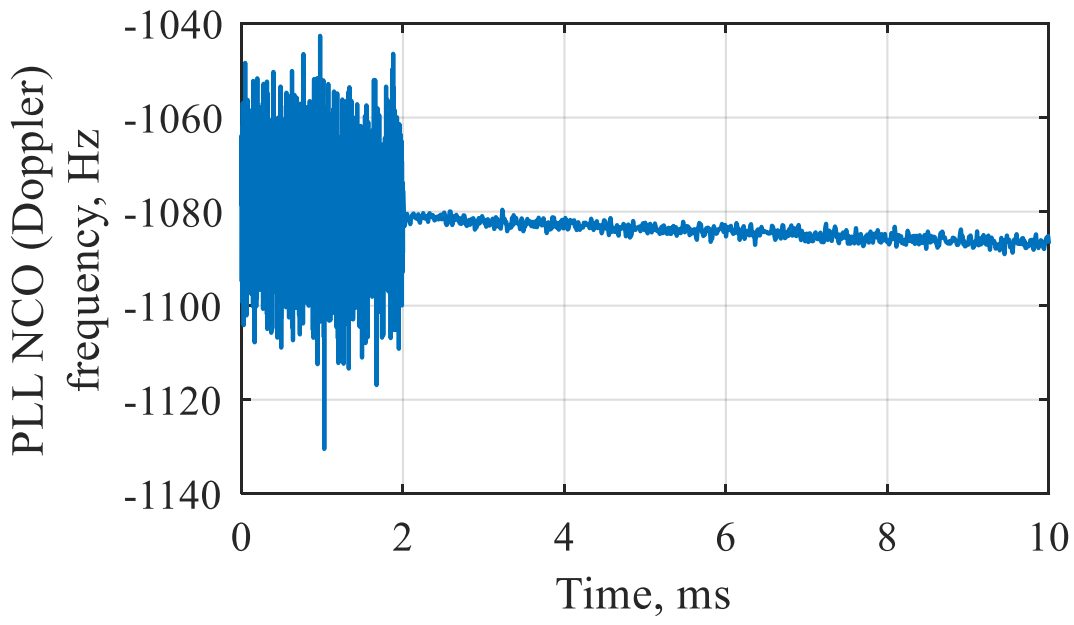

Figure 16. PLL NCO (Doppler) frequency

The data in Figure 17 shows the deviation of the DLL NCO frequency from the nominal value of 1.023 MHz. The averaged version was obtained using a moving average with a window of 64 samples. It can be seen that the noise level consistently decreases as the correlator channel spacing is reduced.

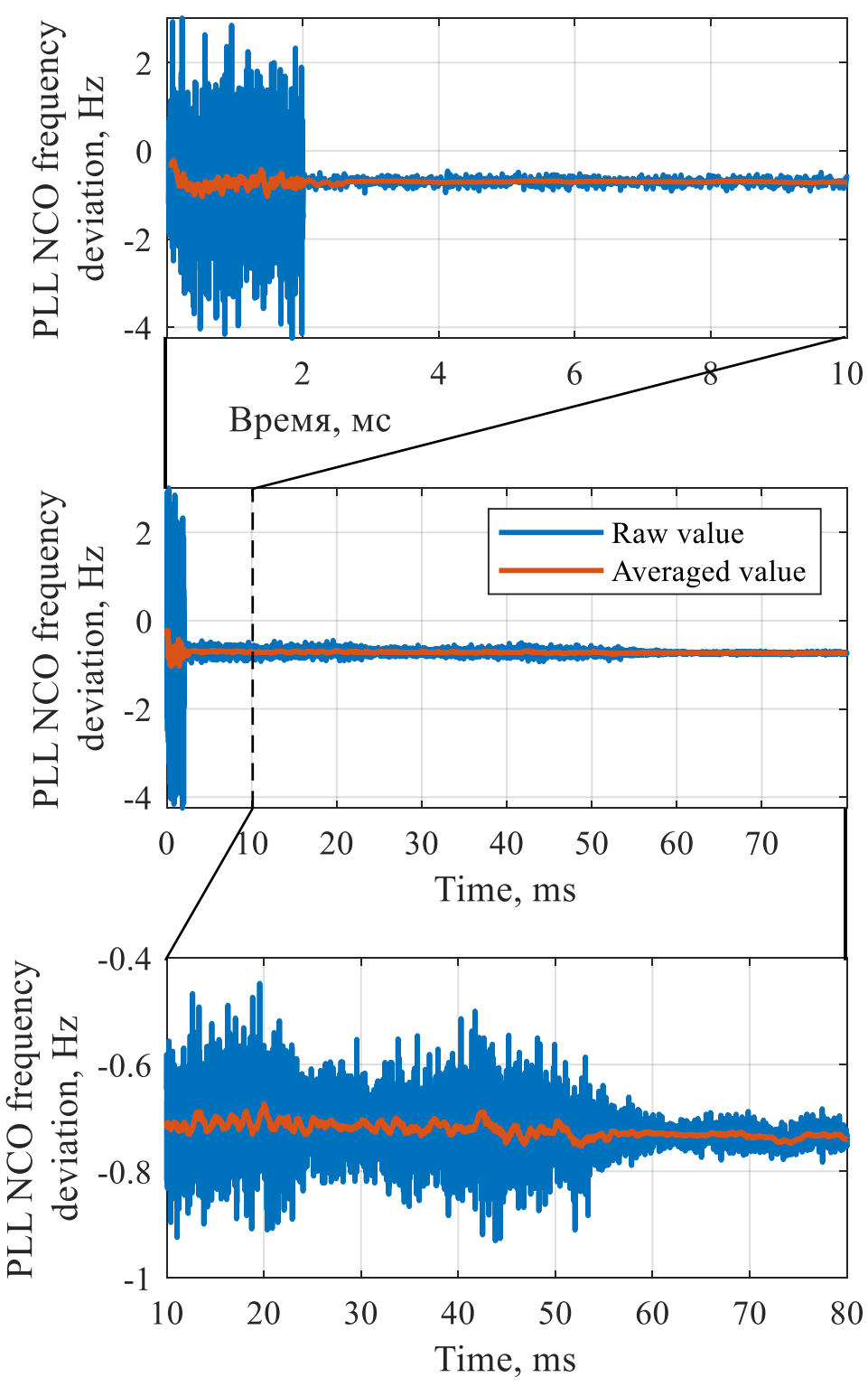

Figure 17. Frequency deviation of the DLL NCO

## Conclusion

Tracking accuracy can be improved by applying extended integration intervals and reduced spacing between the early and late correlator channels. When implementing these methods, several aspects described above must be considered. One key factor is the selection of an appropriate tracking loop filter. Simulation results show that a second-order filter is close to optimal under low-dynamic conditions. In the case of highly dynamic conditions, preference should be given to a third-order filter, but its stability must be carefully monitored.

For classic GPS and GLONASS signals, using extended integration intervals requires synchronization with the navigation data stream. Moreover, the moment of transition from 1 ms to a longer interval can cause a transient process in the PLL /DLL and even a loss of lock. Reduced spacing between correlator channels improves tracking accuracy in the DLL; however, switching can cause a transient process lasting several seconds due to the narrow bandwidth of the filter.